\documentclass[structabstract]{aa}  
\usepackage{graphicx}
\usepackage{txfonts}
\usepackage[round]{natbib}

\begin{document}

\title{Millimeter interferometric observations of\\ FU\,Orionis-type
  objects in Cygnus}

\author{\'A. K\'osp\'al
          \inst{1}}
        \institute{Leiden Observatory, Leiden University, PO Box 9513,
          2300 RA Leiden, The Netherlands\\
              \email{kospal@strw.leidenuniv.nl}
}

 \date{Received date; accepted date}

\abstract
{FU\,Orionis-type objects (FUors) are low-mass young eruptive stars
  that probably represent an evolutionary phase characterized by
  episodic periods of increased accretion rate from the circumstellar
  disk to the star. Theory predicts that a circumstellar envelope, the
  source of continuous mass infall onto the disk, is necessary for
  triggering such accretion bursts.}
{We intend to study the spatial and velocity structure of
  circumstellar envelopes around FUors by means of molecular line
  observations at millimeter wavelengths. We target three prototypical
  FUors as well as an object possibly in a pre-outburst state.}
{We present archival PdBI interferometric observations of the J=1--0
  line of $^{13}$CO at 110.2\,GHz. For three of our targets, these
  represent the first millimeter interferometric observations. The
  data allow the study of the molecular environment of the objects on
  a spatial resolution of a thousand AU and a velocity resolution of
  0.2\,km\,s$^{-1}$.}
{Strong, narrow $^{13}$CO(1--0) line emission is detected from all
  sources. The emission is spatially resolved in all cases, with
  deconvolved sizes of a few thousand AUs. For V1057\,Cyg and
  V1331\,Cyg, the emitting area is rather compact, suggesting that the
  origin of the emission is a circumstellar envelope surrounding the
  central star. For V1735\,Cyg, the $^{13}$CO emission is offset from
  the stellar position, indicating that the source of this emission
  may be a small foreground cloud, also responsible for the high
  reddening of the central star. The $^{13}$CO emission towards
  V1515\,Cyg is the most extended in the sample, and apparently
  coincides with the ring-like optical reflection nebula associated
  with V1515\,Cyg.}
{We suggest that millimeter interferometric observations are
  indispensable for a complete understanding of the circumstellar
  environment of FUors. Any theory of the FUor phenomenon that
  interprets the geometry of the circumstellar structure and its
  evolution using single beam measurements must be checked and
  compared to interferometric observations in the future.}


\maketitle


\section{Introduction}

FU\,Orionis-type objects, or shortly FUors, constitute a small group
of young stars characterized by large outbursts in visible light,
attributed to highly enhanced accretion \citep{hk96}. During these
outbursts, accretion rates from the circumstellar disk to the star are
in the order of 10$^{-4}$\,M$_{\odot}$/yr, three orders of magnitude
higher than in quiescence. Enhanced accretion is often accompanied by
enhanced mass loss: most FUors have optical jets, molecular outflows,
and optically visible ring-like structures that are sometimes
explained by expanding shells thrown off during previous outbursts.

During a single, century-long outburst, as much as 0.01 M$_{\odot}$ of
material can be dumped onto the stellar surface. Thus, the inner disk
needs to be replenished after each outburst, possibly by material from
the infalling envelope. Recent theoretical studies show that the
continuous infall from the envelope is also necessary to trigger the
outbursts \citep{vorobyov2006}. After many outbursts, the envelope
vanishes, and the object finally enters a state of permanently low
accretion. This general paradigm of the evolution of young, low-mass
stars was invoked by \citet{quanz2007} to explain the observed
diversity of FUors: while some objects are still deeply embedded
(e.g.~L1551\,IRS\,5, V1735\,Cyg), others have already cleared away
part of their envelopes (e.g.~V1057\,Cyg, V1515\,Cyg). Statistics show
that probably all low-mass young stars undergo FUor-like phases during
their evolution, implying that FUors might be the clue objects to
study envelope evolution and dispersal.

Interferometric observations of molecular line emission have been
successfully used to probe small-scale structure of molecular material
around young stellar objects. For example, for a sample of low-mass
YSOs in Taurus \citep{hogerheijde1998}, Serpens
\citep{hogerheijde1999}, and Ophiuchus \citep{vankempen2009}, it was
found that most YSOs are surrounded by compact envelopes of up to a
few thousand AU in radius, well traced by $^{13}$CO and
C$^{18}$O. Condensations, inhomogeneities in the envelopes can be seen
in HCO$^+$ and $^{13}$CO. The HCO$^+$ and HCN molecules trace the
walls of the outflow, while SiO and SO emission originates from
shocked material in the outflow. These observations provided detailed
kinematical picture of the envelopes (for molecular studies of some
individual young stellar objects, see e.g.~\citealt{jorgensen2004,
  matthews2006, brinch2009}).

Despite being ``accretion laboratories'', there are relatively few
molecular gas observations published for FUors. Single-dish
observations often show the presence of strong CO emission towards
FUors, and line profiles sometimes indicate the presence of molecular
outflows (e.g.~\citealt{hk96}, and references therein). However, due
to the large single dish beam, it is not possible to study the spatial
distribution of the emission using these data. Out of the 9 FUors
listed in \citet{hk96}, millimeter interferometric observations are
only available for L1551\,IRS\,5 \citep[e.g.][]{momose1998}.

In this paper, we present interferometric observations of the
molecular emission from four FUor-type objects in Cygnus:
\object{V1057\,Cyg}, \object{V1331\,Cyg}, \object{V1515\,Cyg}, and
\object{V1735\,Cyg}. With the exception of V1331\,Cyg
\citep{levreault1988, mcmuldroch1993}, these data represent the first
millimeter interferometric observations for our targets. We use the
$^{13}$CO(1--0) line emission to perform a high spatial and spectral
resolution study of the gaseous material around FUors. We analyze the
spatial and kinematic structure of the $^{13}$CO gas, and check
whether the emission can be associated with the circumstellar
envelopes. Our results can be compared with those obtained for normal
YSOs. Such a comparison may also contribute to the on-going debate
whether all young stars undergo an eruptive phase in their early
evolution or FUors are atypical objects.


\section{Observations}

We reduced unpublished $^{13}$CO observations of our targets obtained
with the Plateau de Bure Interferometer (PdBI) on May 30 and 31, 1993
(program ID: C057, PI: D.~Fiebig). The observations were carried out
in snapshot mode (about 1 hour on-source correlation time per object),
using four antennas (4D1 configuration), and the baselines ranged from
24 to 64\,m. The receiver was tuned to the $^{13}$CO(1--0) line at
110.2\,GHz (lower sideband), the channel spacing was 78\,kHz
(0.21\,km/s). At this wavelength, the single dish HPBW is
45$''$. Bright quasars (3C454.3, 2005+403, 2037+511, 2200+420,
0648--165, and 0727--115) were observed to enable RF bandpass, phase,
and amplitude calibration. The phase center was 20$^{\rm h}$58$^{\rm
  m}$53$\fs$7 +44$^{\circ}$15$'$27$\farcs$9 for V1057\,Cyg, 21$^{\rm
  h}$01$^{\rm m}$09$\fs$3 +50$^{\circ}$21$'$42$\farcs$2 for
V1331\,Cyg, 20$^{\rm h}$23$^{\rm m}$47$\fs$9
+42$^{\circ}$12$'$24$\farcs$8 for V1515\,Cyg, and 21$^{\rm h}$47$^{\rm
  m}$20$\fs$7 +47$^{\circ}$32$'$04$\farcs$1 for V1735\,Cyg.

Data reduction was done in the standard way with CLIC, a GILDAS-based
application written especially for the reduction of PdBI
data\footnote{http://www.iram.fr/IRAMFR/GILDAS}. The rms phase noise
was less than 18$^{\circ}$ for all observations. We estimate a flux
calibration accuracy of $\approx$15\%. MAPPING was used to create
naturally weighted dirty maps in an area of 64$''\,{\times}\,$64$''$,
which were then deconvolved using the Clark CLEAN method. The
synthesized beam is approximately 7$''\,{\times}\,$6$''$ while the rms
noise in the channel maps is $\approx$0.15 Jy/beam. For the exact
phase centers and beam parameters, see the log of observations in
Table~\ref{tab:log}.

\section{Results}

\begin{figure}
\begin{center}
\includegraphics[width=39.2mm,angle=90]{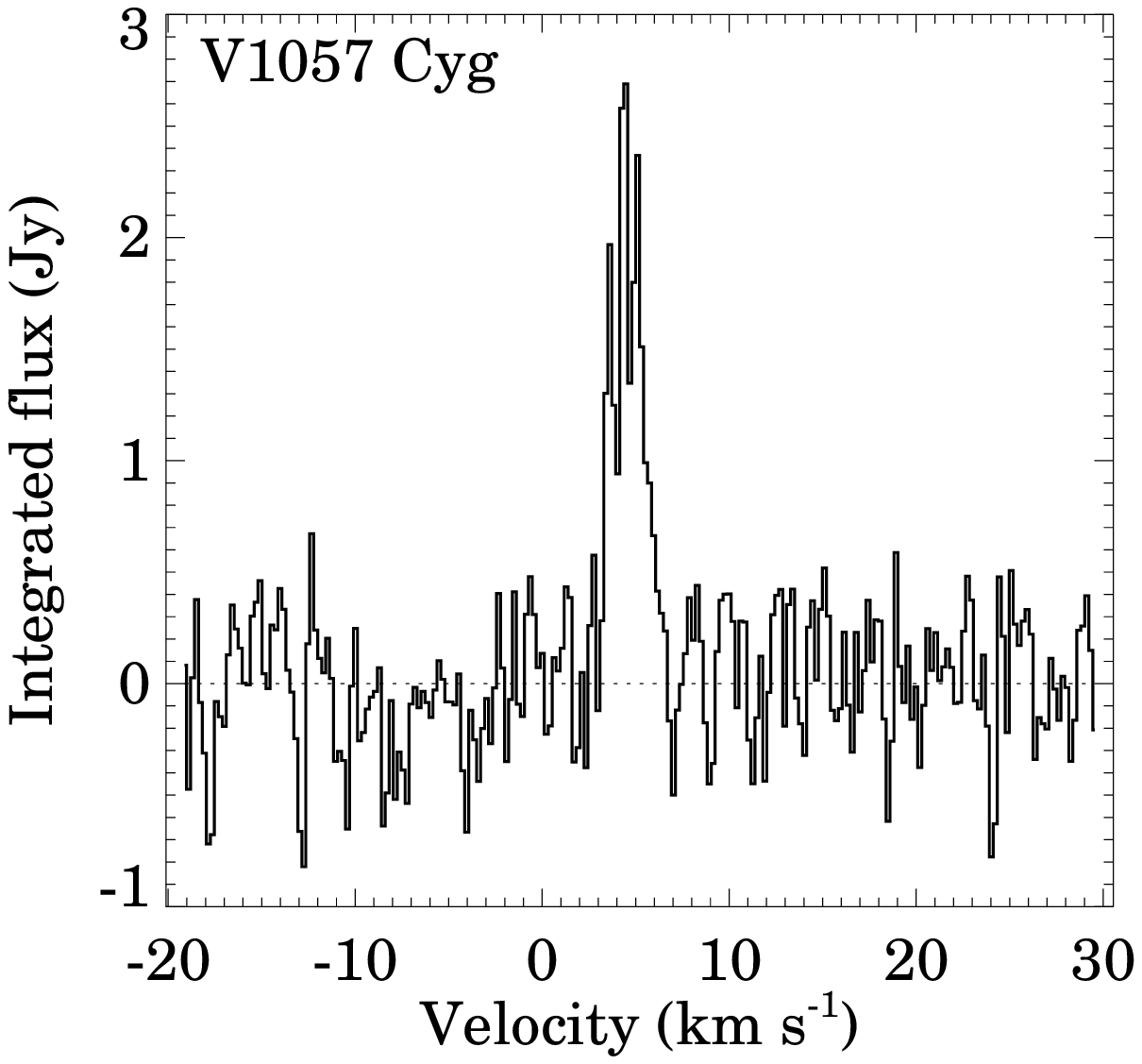}
\includegraphics[width=39.2mm,angle=90]{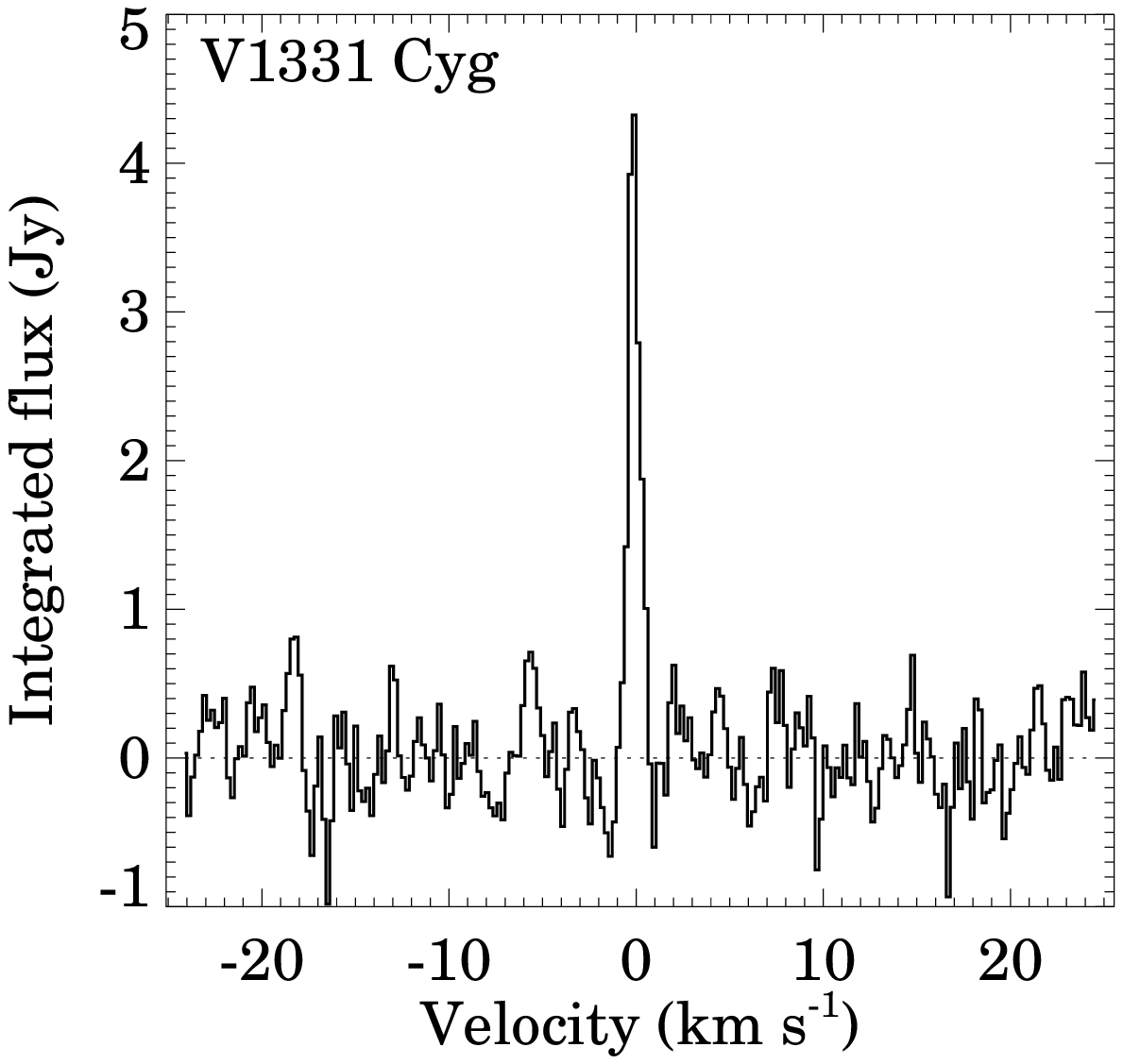}
\includegraphics[width=39.2mm,angle=90]{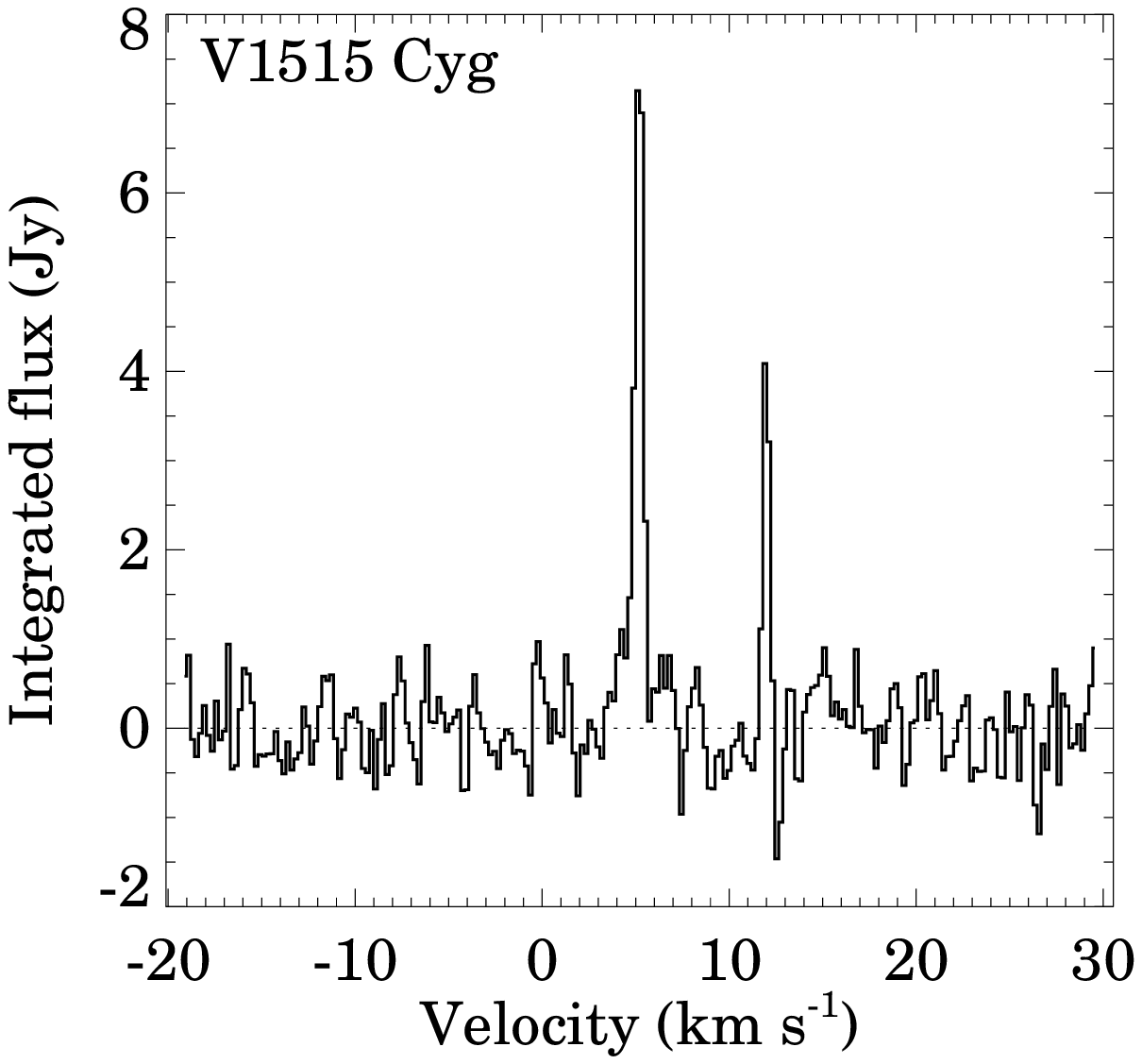}
\includegraphics[width=39.2mm,angle=90]{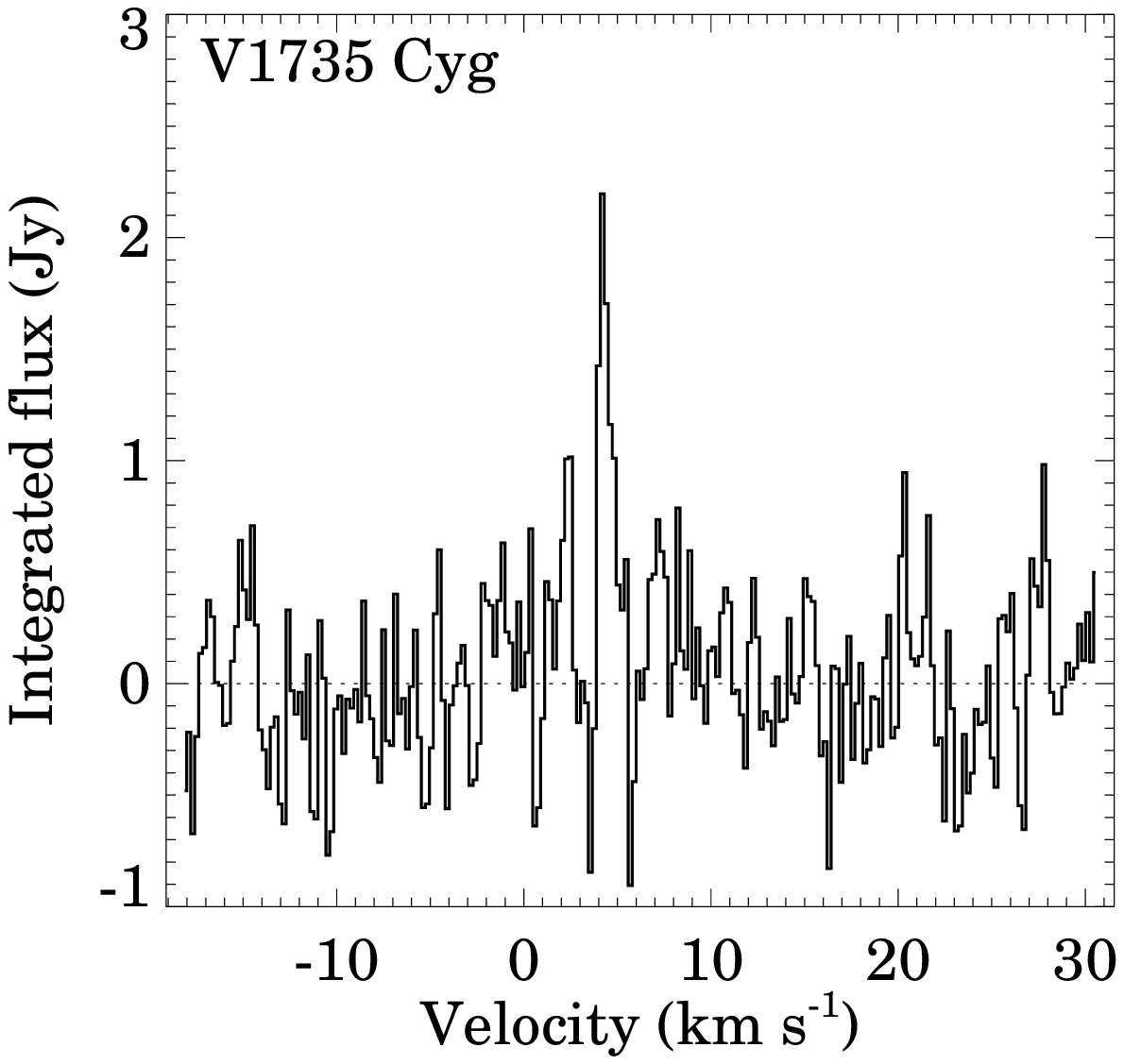}
\caption{$^{13}$CO(1--0) spectra of our targets obtained with the
  PdBI. Fluxes were integrated over an area of $64''\times 64''$
  centered on the phase center.\label{fig:spec}}
\end{center}
\end{figure}

\begin{figure}
\begin{center}
\includegraphics[width=39.2mm,angle=90]{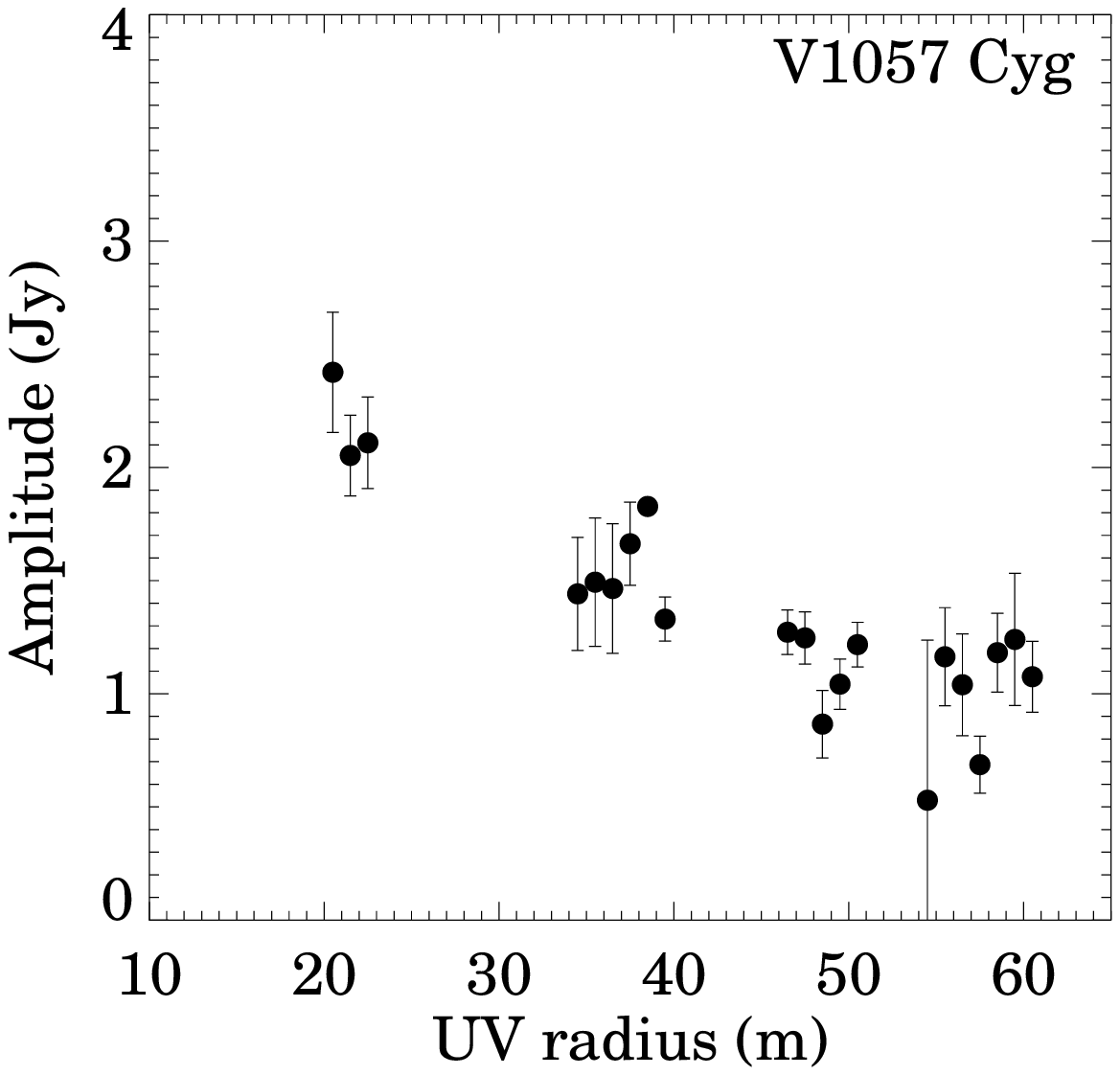}
\includegraphics[width=39.2mm,angle=90]{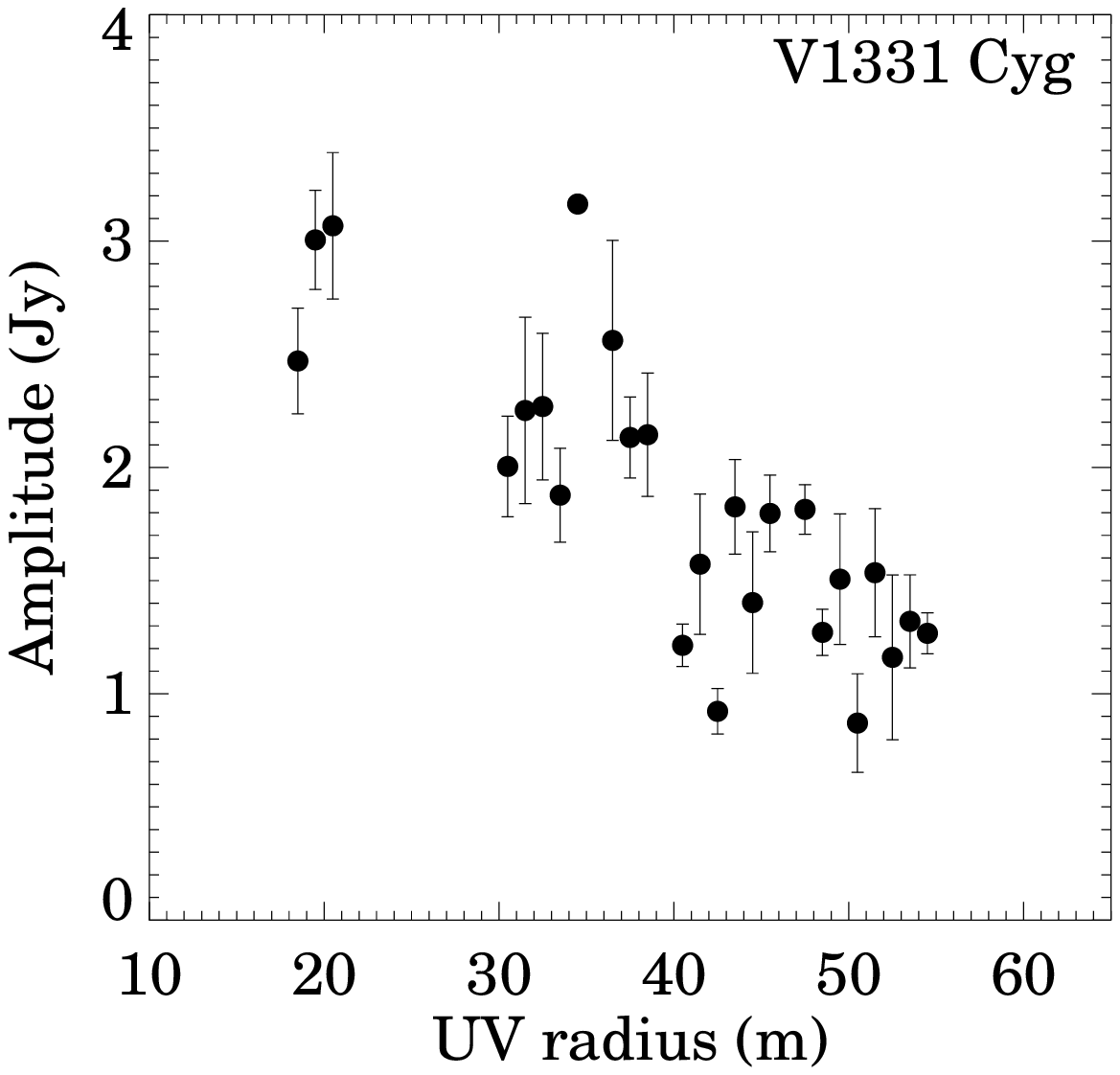}
\includegraphics[width=39.2mm,angle=90]{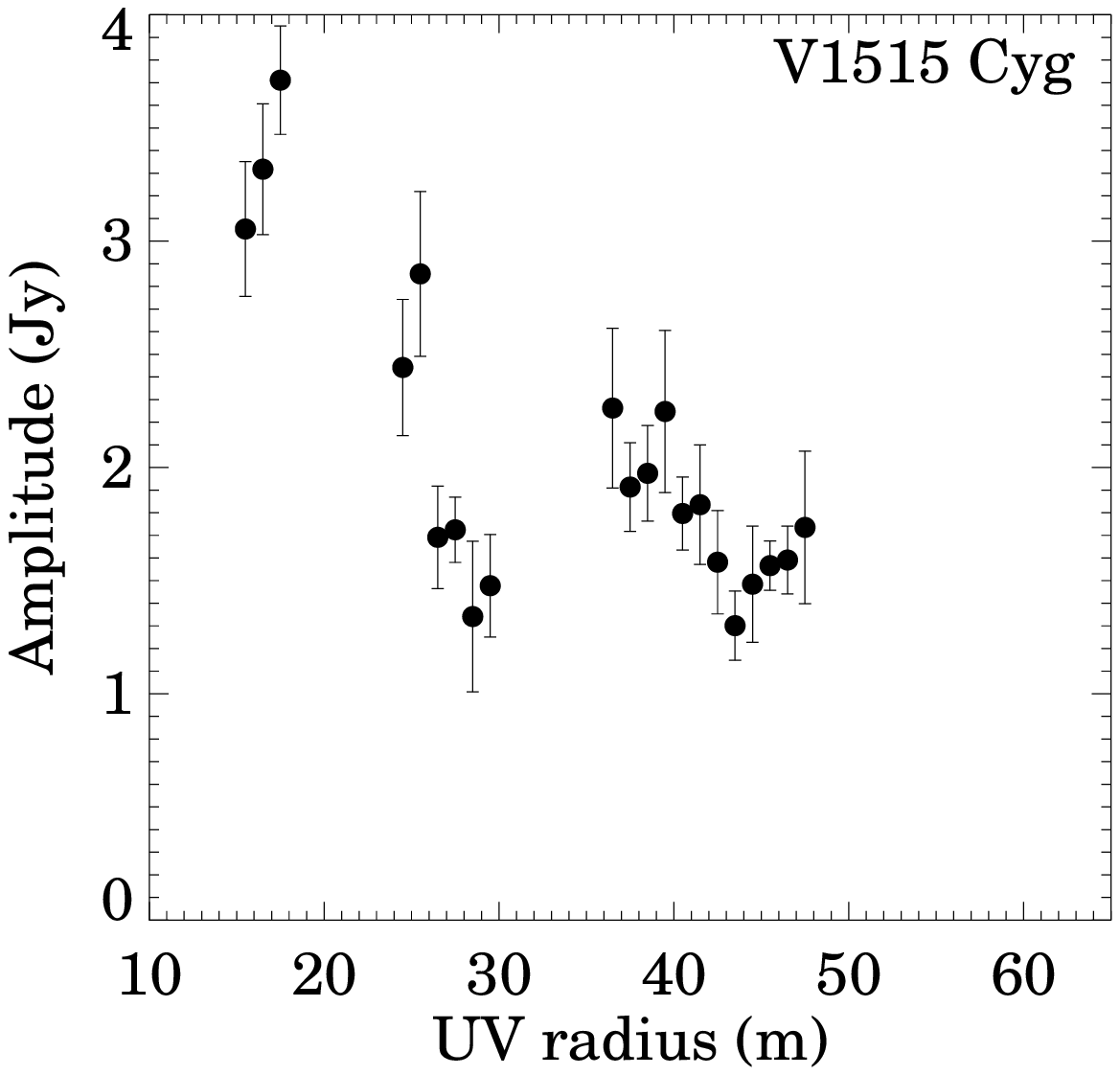}
\includegraphics[width=39.2mm,angle=90]{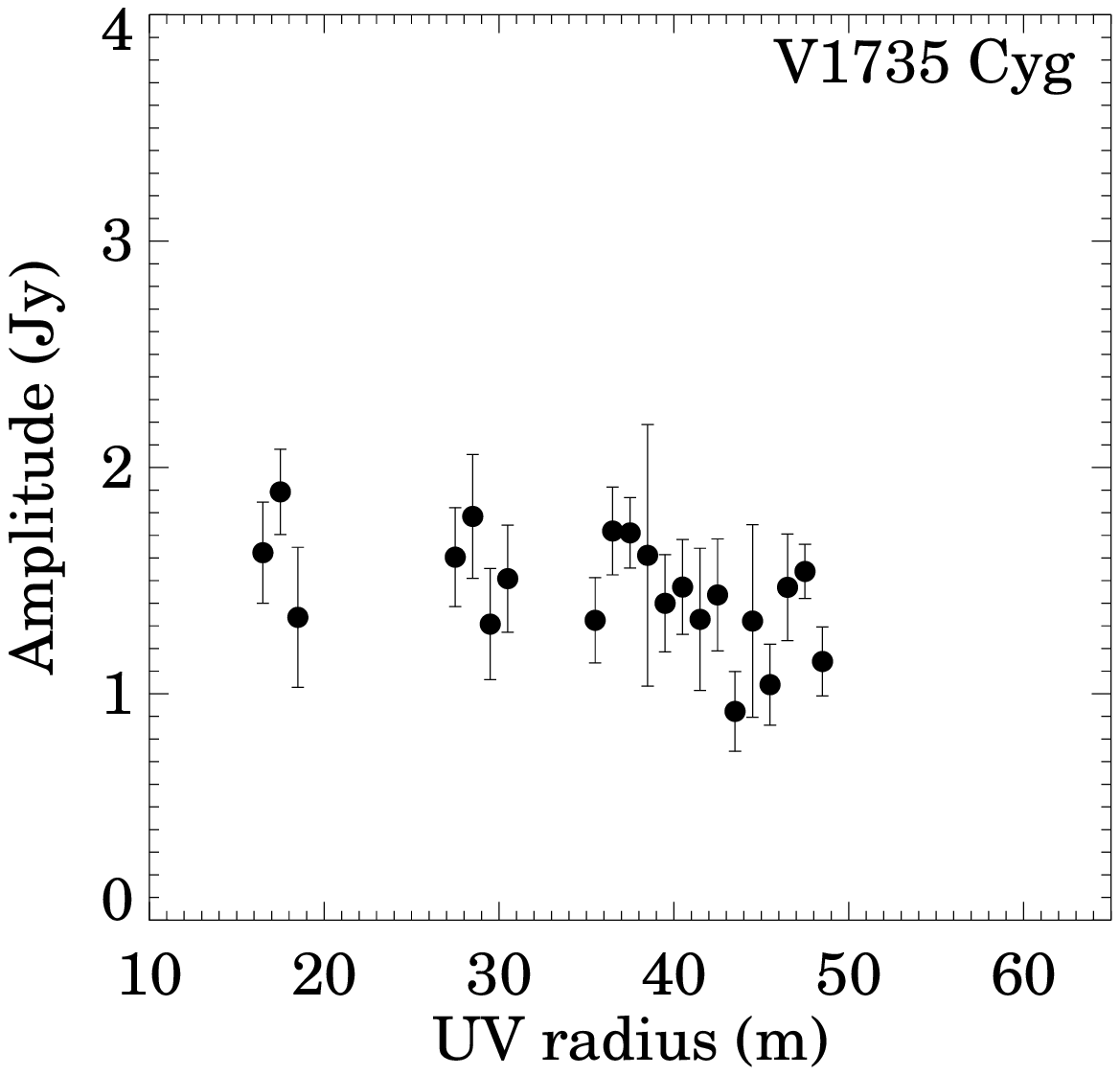}
\caption{Visibility amplitudes as a function of $uv$ radius. Data were
  binned in 1\,m-wide bins and the error bars indicate
  the dispersion of data points within one bin.\label{fig:uv}}
\end{center}
\end{figure}

The $^{13}$CO(1--0) spectra of our sources, integrated over the whole
64$''\,{\times}\,$64$''$ maps, are plotted in
Fig.~\ref{fig:spec}. Strong $^{13}$CO(1--0) emission was detected
towards all sources with high signal-to-noise ratio. In all cases, we
found an emission line coinciding with the systemic velocity of each
source (from optical and near-IR spectra, \citealt{herbig1977},
\citealt{kenyon1989}, and \citealt{chavarria1979} determined
heliocentric radial velocities of $-$14$\,{\pm}\,$2,
$-$15$\,{\pm}\,$2, $-$12$\,{\pm}\,$2, and
$-$10$\,{\pm}\,$2\,km\,s$^{-1}$, coresponding to LSR velocities of
1.9, 0.6, 5.2, and 3.9\,km\,s$^{-1}$ for V1057\,Cyg, V1331\,Cyg,
V1515\,Cyg, and V1735\,Cyg, respectively). We collapsed a few channels
centered on the peak emission, and calculated visibility amplitudes as
a function of $uv$ radius (Fig.~\ref{fig:uv}), and produced
velocity-integrated maps (Fig.~\ref{fig:beam}). The LSR velocities
where the emission peaks are listed in
Table~\ref{tab:log}. Velocity-integrated line fluxes calculated for
the emission areas visible in Fig.~\ref{fig:beam} are also listed in
Table~\ref{tab:log}.

\begin{table*}
\begin{center}
\begin{tabular}{c@{}c@{}ccccccccccc}
\hline
           &          &             & \multicolumn{2}{c}{Synthesized beam}  & & \multicolumn{2}{c}{Peak position}       & & \multicolumn{2}{c}{Fitted Gaussian}   &                &                    \\
\cline{4-5} \cline{7-8} \cline{10-11}\\
Name       & Distance & Date        & FWHM                   & P.A.         & & $\alpha_{2000}$ & $\delta_{2000}$       & & FWHM                   & P.A.         & $v_{\rm LSR}$  & Flux               \\
           & (pc)     &             & ($''\,{\times}\,''$)   & ($^{\circ}$) & & (h:m:s)         & ($^{\circ}$:$'$:$''$) & & ($''\,{\times}\,''$)   & ($^{\circ}$) & (km\,s$^{-1}$) & (Jy\,km\,s$^{-1}$) \\
\hline
V1057 Cyg  & 600      & 1993-May-31 & 7.02$\,{\times}\,$5.16 & 2            & & 20:58:53.8      & +44:15:28.7           & & 9.6$\,{\times}\,$5.2   & 177          & 4.6            & 4.1                \\
V1331 Cyg  & 550      & 1993-May-31 & 7.30$\,{\times}\,$5.97 & 147          & & 21:01:09.2      & +50:21:44.2           & & 9.7$\,{\times}\,$6.5   & 148          & $-$0.1         & 2.5                \\
V1515 Cyg  & 1000     & 1993-May-30 & 7.43$\,{\times}\,$6.90 & 148          & & 20:23:48.0      & +42:12:30.0           & & 21.9$\,{\times}\,$10.6 & 166          & 5.1            & 5.1                \\
V1735 Cyg  & 900      & 1993-May-30 & 8.01$\,{\times}\,$6.53 & 133          & & 21:47:20.5      & +47:32:06.3           & & 9.4$\,{\times}\,$9.4   & 78           & 4.2            & 1.5                \\
\hline
\end{tabular}
\caption{Summary of PdBI observations. Distances are from
  \citet{sw2001}. The FWHM of the fitted Gaussians are deconvolved
  sizes. \label{tab:log}}
\end{center}
\end{table*}

\subsection{Line profiles}

The line profile of V1057\,Cyg is the broadest in our sample
(FWHM$\approx$2.1\,km\,s$^{-1}$), and is centered at
4.6\,km\,s$^{-1}$. This is consistent with the velocity seen in single
dish $^{12}$CO(1--0), $^{12}$CO(2--1), and $^{13}$CO(1--0) data by
\citet{bechis1975} and \citet{levreault1988}. V1331\,Cyg shows a
narrow $^{13}$CO(1--0) line
(FWHM$\approx$0.7\,km\,s$^{-1}$). \citet{mcmuldroch1993} presents
Owens Valley interferometric observations of V1331\,Cyg in the same
line, as well as in other transitions and other isotopes of CO. The
velocities of these lines are consistent with ours. V1515\,Cyg shows
two separate, narrow emission components
(FWHM$\approx$0.6\,km\,s$^{-1}$) at 5.1\,km\,s$^{-1}$ and at
11.9\,km\,s$^{-1}$. Both of these components are also visible in
single dish $^{12}$CO(3--2) and $^{13}$CO(2--1) data from
\citet{evans1994}. The spectrum of V1735\,Cyg shows a narrow,
single-peaked $^{13}$CO(1--0) line
(FWHM$\approx$0.7\,km\,s$^{-1}$). The shape and position of this line
coincides well with that of the $^{13}$CO(2--1) line detected in
single dish data by \citet{evans1994}. The $^{12}$CO(2--1) line from
the same paper shows self-absorption at this velocity, and broad line
wings indicating outflow activity. We plotted position-velocity
diagrams along different angles through our sources, but except for
V1515\,Cyg (which will be discussed later), we found no significant
velocity gradients.

\subsection{Spatial extent}

\begin{figure}
\begin{center}
\includegraphics[width=39.2mm,angle=90]{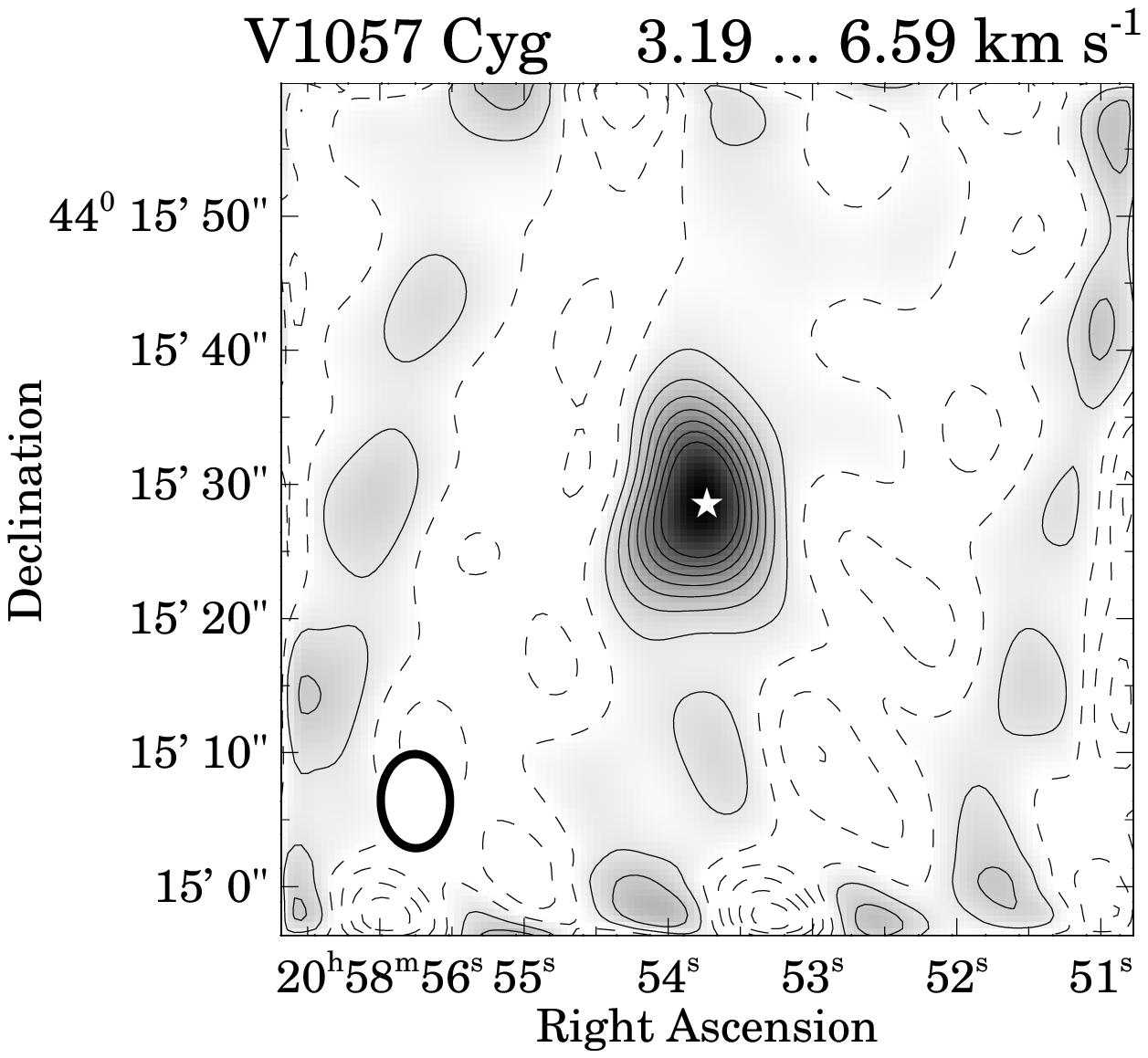}
\includegraphics[width=39.2mm,angle=90]{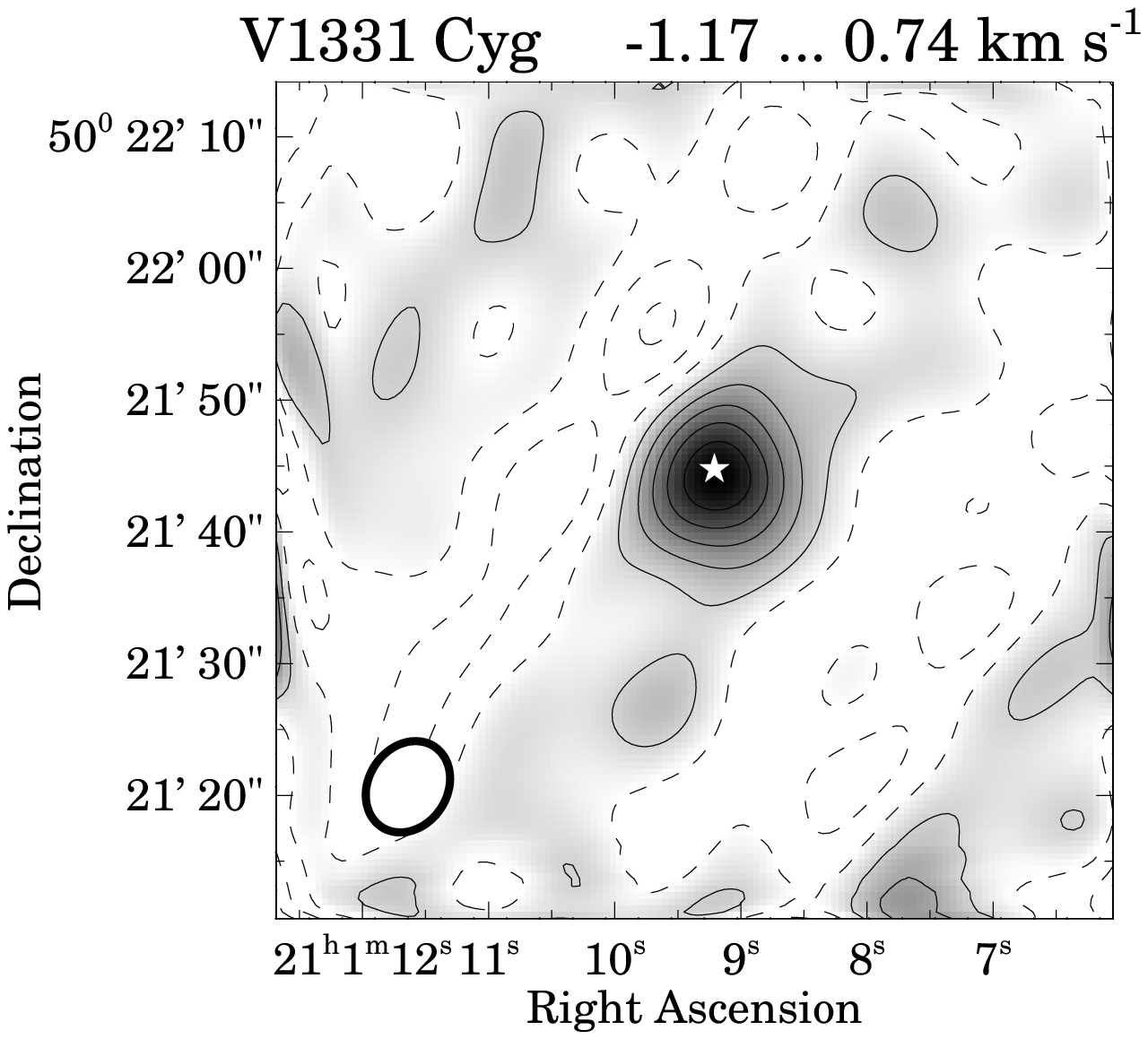}
\includegraphics[width=39.2mm,angle=90]{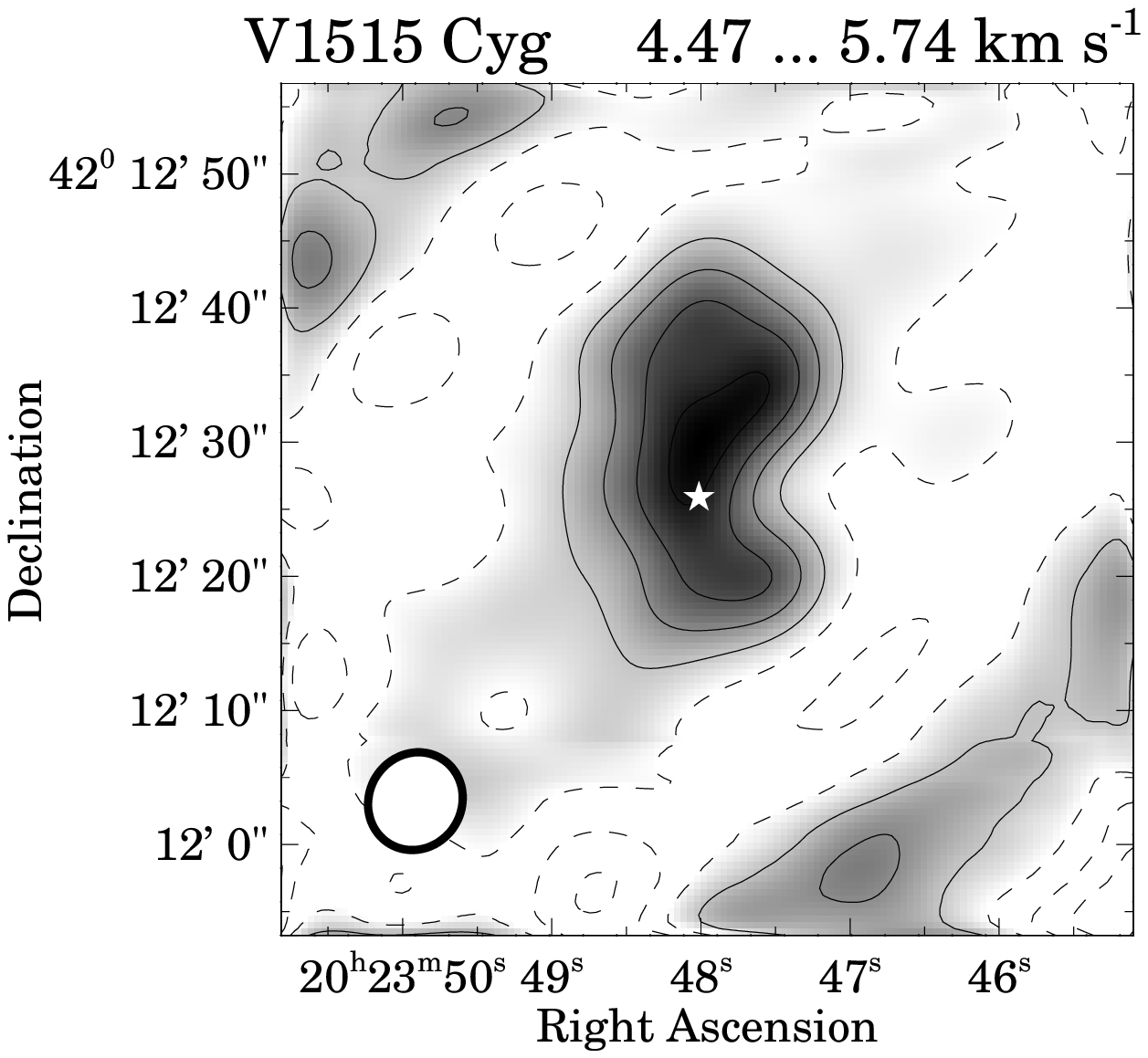}
\includegraphics[width=39.2mm,angle=90]{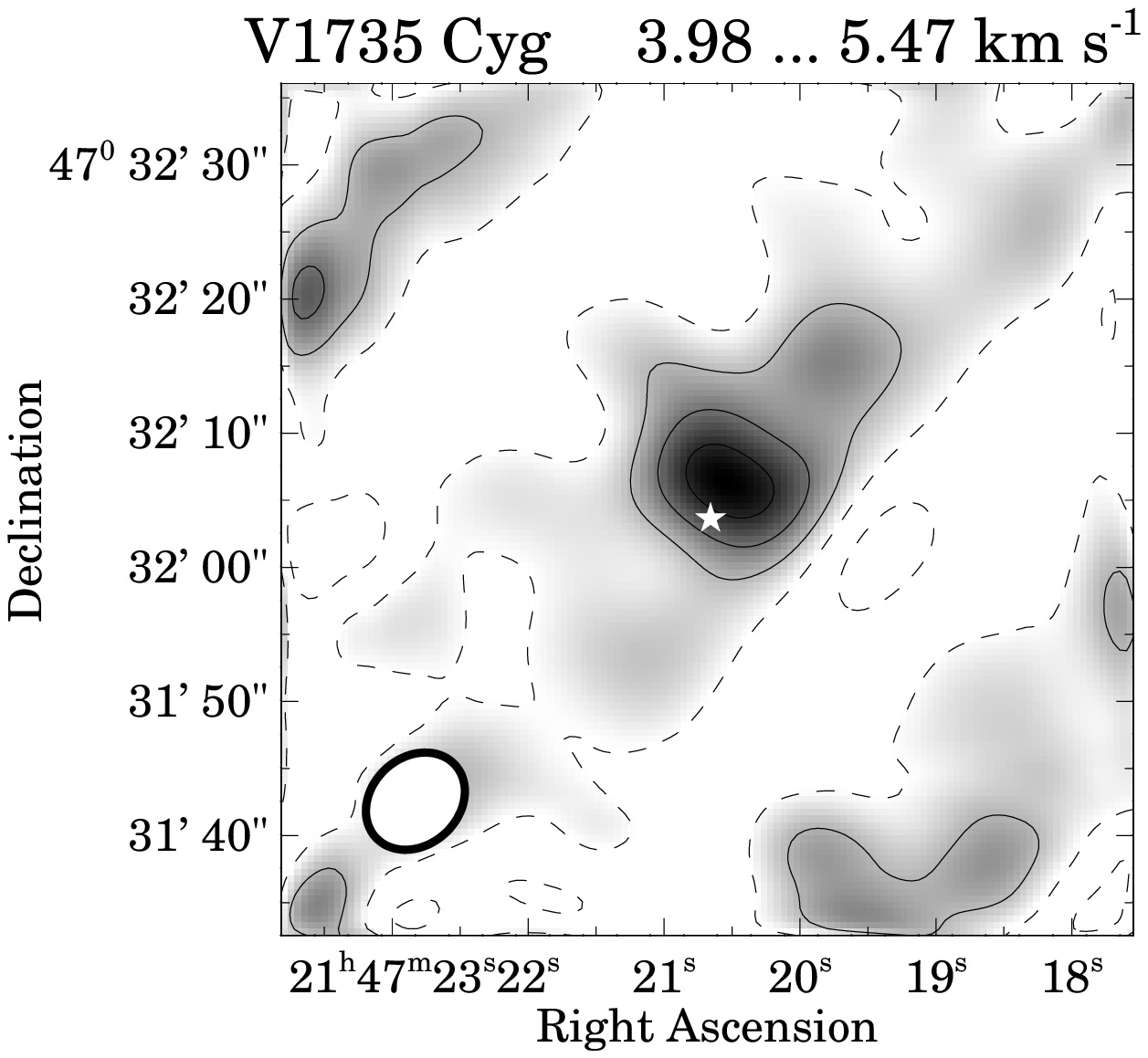}
\caption{$^{13}$CO(1--0) maps of our targets, integrated in the
  velocity ranges indicated above the images. White asterisks indicate
  optical positions from SIMBAD. Beam sizes are
  $\approx$6$\times$7$''$. The noise level is
  $\sigma$=0.10\,Jy\,km\,s$^{-1}$; the solid contours are 2$\sigma$,
  4$\sigma$, 6$\sigma$, ...; the dashed contours are 0, $-$2$\sigma$,
  $-$4$\sigma$.\label{fig:beam}}
\end{center}
\end{figure}

In Fig.~\ref{fig:uv} we plotted the visibility amplitudes as a
function of $uv$ radius. Since these are snapshot observations, the
$uv$-space is only partially sampled; there are ranges of $uv$ radii
where no visibility information is available. However, even with this
coarse sampling, the graphs show a decline of amplitude with $uv$
radius, indicating that our sources are spatially resolved. Simple 2D
Gaussian fits to the maps presented in Fig.~\ref{fig:beam} also
indicate that our targets are spatially resolved. The deconvolved
sizes and PAs of the Gaussian fits are listed in
Table~\ref{tab:log}. Gaussian models fitted to the visibilities give
sizes within $\pm$2$''$ and PAs within $\pm$10$^{\circ}$ to those
determined from the Gaussian fits to the images. Thus, we consider
these numbers as representative uncertainties of the values given in
Table~\ref{tab:log}. We emphasize that these fits are not physical
envelope models, and thus the obtained parameters should be regarded
only as rough quantitative estimates of the size of the emitting
region.

V1057\,Cyg and V1331\,Cyg are the most compact sources in our sample,
with FWHM of about 10$''\,{\times}\,$5$''$. Moreover, in both cases,
the emission in centered precisely on the optical position of the
stars. V1515\,Cyg shows the most extended emission in our sample. The
stellar position seems to be located at the southern tip of a bright,
slightly curved, elongated filament, but extended emission can be seen
both to the south and to the north of the star ($\approx$20$''$
across). A map produced for the channels around the emission line at
11.9\,km\,s$^{-1}$ reveals that this separate velocity component comes
from a compact area $\approx$13$''$ to the northwest
(P.A. 330$^{\circ}$) of the optical source. The bulk of the emission
in the V1735\,Cyg region comes from an area with a FWHM of
$\approx$9$''$ centered 2\farcs9 to the northwest with respect to the
optical position of the star. Moreover, there is also some tentative
extended emission towards the northwest about 14$''$ from the stellar
position.

\subsection{Gas masses}

Following \citet{scoville1986}, the total H$_2$ mass can be calculated
from the observed $^{13}$CO(1--0) line fluxes by
\[
M_{H_2}=2.39 \times 10^{-9} \times \frac{(T_X+0.89)}{e^{-5.31/T_X}}
\frac{\tau_{^{13}\rm CO}}{1-e^{-\tau}} \frac{D^2_{\rm kpc}}{X(^{13}\rm
  CO)} \int S_{\nu}dv \, M_{\odot},
\]
where $T_X$ is the excitation temperature, $\tau$ is the optical
depth, $D$ is the distance of the source, $X(^{13}\rm CO)$ is the
molecular abundance relative to H$_2$, and $\int S_{\nu}dv$ is the
velocity-integrated line flux in units of Jy\,km\,s$^{-1}$. Using
$T_X$=50\,K (as determined from dust continuum observations by
\citealt{sw2001}), assuming the lines to be optically thin
($\tau{\ll}$1), and taking $X(^{13}\rm CO)$=1.6$\times$10$^{-6}$
\citep{langer1993}, the total gas masses are 0.12, 0.06, 0.43, and
0.10\,M$_{\odot}$ for V1057\,Cyg, V1331\,Cyg, V1515\,Cyg, and
V1735\,Cyg, respectively. Of course these numbers should be considered
lower limits if the lines are actually optically thick, or if part of
the $^{13}$CO emission is resolved out by the interferometer. However,
these values are in good agreement with total masses derived from
single-dish 850$\,\mu$m dust continuum maps of comparable size by
\citet{sw2001}, which were 0.10, 0.13, 0.15, and 0.42\,M$_{\odot}$ for
V1057\,Cyg, V1331\,Cyg, V1515\,Cyg, and V1735\,Cyg, respectively. This
suggests that our interferometric observations probably recover most
of the $^{13}$CO emission.

\subsection{2.7\,mm continuum}

Although the main focus of our data analysis is the study of the
$^{13}$CO emission, the PdBI observations could also be used to look
for 2.7\,mm continuum emission by excluding channels around the lines
and collapsing the remaining channels. The resulting continuum maps
have a typical rms noise of 2\,mJy/beam. We detected 2.7\,mm continuum
emission only for one of our targets, V1331\,Cyg. This is not
surprising because, based on the observations of \citet{sw2001}, out
of our four targets, V1331\,Cyg is the brightest also at 1.3\,mm and
at 850$\,\mu$m. The 2.7\,mm emission we detected is not resolved, its
position is consistent with the optical stellar position, and the
measured total flux is 12$\pm$2\,mJy. Assuming optically thin
emission, dust opacity of $\kappa_{1.3\rm mm}$=0.01\,cm$^2$\,g$^{-1}$
\citep{ossenkopf1994}, an emissivity law of
${\kappa}\,{\sim}{\lambda}^{-1}$ and dust temperature of 50\,K
\citep{sw2001}, this flux corresponds to a total (gas+dust) mass of
0.19\,M$_{\odot}$, similar to what we obtained from the line flux of
V1331\,Cyg. Using 6\,mJy as a 3$\sigma$ upper limit for the 2.7\,mm
continuum flux of the other sources, we can give upper limits of 0.11,
0.32, and 0.26\,M$_{\odot}$ for V1057\,Cyg, V1515\,Cyg, and
V1735\,Cyg, respectively.

\section{Discussion}

In the following subsections, we discuss what our interferometric
$^{13}$CO(1--0) data can add to what we already know about the
circumstellar environment of our targets based on optical images,
sub-millimeter continuum maps, CO line observations, and SED analysis
from the literature. We will also discuss whether the detected
$^{13}$CO emission can be associated with the envelopes.

\subsection{V1057\,Cyg}

After its outburst in 1969-70, an eccentric ring-like nebulosity of
1$'\,{\times}\,$1$\farcm$5 appeared around V1057\,Cyg
\citep{herbig1977}. The monitoring of this ring in the subsequent
years indicated that it is fading together with the central star
without any changes in the structure. This implies that the ring is a
reflection nebula: a pre-existing structure illuminated by the
outbursting star, and not material ejected from the central source by
the outburst. From the line wings of single dish $^{12}$CO data,
\citet{levreault1988} and \citet{evans1994} concluded the presence of
a molecular outflow on a similar spatial scale. The surroundings of
V1057\,Cyg was mapped by \citet{sw2001}, whose 850$\,\mu$m continuum
image indicates the presence of a rather compact but resolved
($\approx$4$''$) source coinciding with the star and a fainter,
north-south oriented filament.

In our $^{13}$CO(1--0) data we do not detect emission either from the
ring-like reflection nebula of from the north-south filament. However,
our interferometric observations definitely resolve the central
source: the deconvolved Gaussian FWHM at the distance of V1057\,Cyg
corresponds to about 5800$\,{\times}\,$3100\,AU. The lack of
observable line wings, the relatively narrow line profile, and the
compact spatial appearance of the emission points to quiescent gas,
probably situated in the outer parts of a circumstellar envelope.

The infrared SED of the source \citep[e.g.][]{abraham2004} suggests a
continuous temperature distribution in the system. This supports our
conclusion that there is an envelope associated with and heated by the
central star. The presence and measured size of the envelope is
consistent with the model prediction of 7000\,AU proposed by
\citet{green2006}. They also suggested that the envelope has a rather
large conical cavity which explains the modest far-infrared
excess. The relatively low extinction along the line-of-sight and the
detectability of the 10$\,\mu$m silicate emission feature implies a
close to pole-on geometry for the source, in accordance with its
relatively symmetric shape in our interferometric observations. The
circumstellar mass of 0.12\,M$_{\odot}$ derived from our $^{13}$CO
observations, consistently with the dust mass estimate by
\citet{sw2001}, is rather high, significantly exceeding the typical
disk masses of T\,Tauri-type stars.

\subsection{V1331\,Cyg}

V1331\,Cyg is not a FUor in its present state, but -- based on the
similarity of its spectrum to that of V1057\,Cyg prior to maximum
light -- is probably in a pre-outburst state, or between outbursts
\citep{mcmuldroch1993}. Single-dish and interferometric $^{12}$CO and
$^{13}$CO observations by \citet{levreault1988} and
\citet{mcmuldroch1993} revealed a complex circumstellar environment
containing a molecular outflow approximately along the line of sight,
a flattened gaseous envelope of about 6000$\,{\times}\,$4400\,AU in
size, and a gaseous expanding ring of about
41\,000$\,{\times}\,$28\,000\,AU. The latter coincides well with the
large ring-shaped optical reflection nebula seen by \citet{quanz2007}.

Our interferometric $^{13}$CO observations of V1331\,Cyg, consistently
with the similar angular resolution measurements of
\citet{mcmuldroch1993}, reveal a rather compact structure towards the
star with a deconvolved Gaussian FWHM of about
5300$\,{\times}\,$3600\,AU at the distance of V1331\,Cyg. This core is
also visible in continuum emission in the 850$\,\mu$m maps of
\citet{sw2001}, although it is only marginally resolved
($\lesssim$6$''$ or 3300\,AU). We suggest that both the $^{13}$CO gas
emission and the dust thermal emission is originated in a
circumstellar envelope, probably the same flattened structure that was
proposed by \citet{mcmuldroch1993} to explain the nature of the
$^{13}$CO emission. It is probable that the dust in this flattened
envelope scatters the stellar light and gives rise to the inner ring
observed in optical images by \citet{quanz2007}.

The presence of a circumstellar envelope around V1331\,Cyg is also
consistent with the SED, which exhibits significant infrared excess
indicating substantial amounts of circumstellar material
\citep[e.g.][]{abraham2004}. The geometry proposed by
\citet{mcmuldroch1993} and by \citet{quanz2007} suggests that we see
the inner part of the system through a conical cavity filled with a
pole-on molecular outflow. However, we detect $^{13}$CO emission
neither from the outflow nor from the outer, large expanding ring, in
agreement with the interferometric observations of
\citet{mcmuldroch1993}.

\subsection{V1515\,Cyg}

\begin{figure}
\begin{center}
\includegraphics[width=77mm,angle=90]{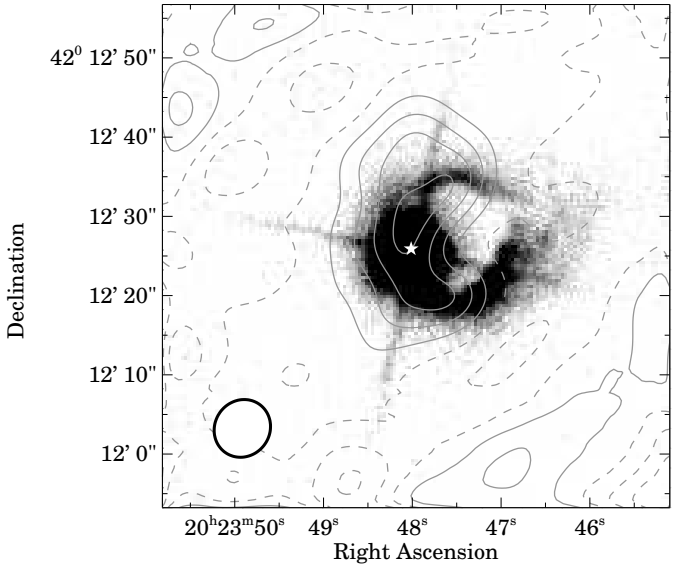}
\caption{The optical and millimeter environment of V1515\,Cyg. The
  grayscale image in the background is an SDSS r-band image; the gray
  contours are the same $^{13}$CO(1--0) contours as in
  Fig.~\ref{fig:beam}. White asterisk indicate the optical stellar
  position.\label{fig:v1515cyg}}
\end{center}
\end{figure}

V1551\,Cyg brightened in optical light slowly during the 1940s and
1950s \citep{herbig1977}. Photographic plates from this time show a
bright, narrow arc of nebulosity extending to the north and west of
the star. Later images from the 1970s to the present day still show
this northern arc, but a brighter nebulosity is also visible to the
south and west, together forming a nearly complete circular ring with
a diameter of $\approx$16$''$ (corresponding to 16\,000\,AU, see also
the Sloan Digital Sky Survey r-band image from 2003 in
Fig.~\ref{fig:v1515cyg}). The fact that the size of this ring did not
change significantly for 60 years but the brightness distribution did,
suggests that it is a reflection nebula similar in nature to that
around V1057\,Cyg.

Our $^{13}$CO(1--0) measurement shows arc-shaped emission, which,
plotted over the optical image in Fig.~\ref{fig:v1515cyg}, clearly
coincides with the ring-shaped reflection nebula. The similar
morphology of the $^{13}$CO emission and the scattered light, as well
as the matching radial velocities of the $^{13}$CO emission and the
star suggest that the detected molecular gas is physically associated
with the FUor. Similarly to V1057\,Cyg, the $^{12}$CO line profile of
V1515\,Cyg in \citet{evans1994} indicate the presence of a molecular
outflow. Our $^{13}$CO(1--0) data, however, show no high-velocity line
wings; the emission line is narrow and thus probably traces quiescent,
not outflowing material. We detect a small
($\lesssim$0.5\,km\,s$^{-1}$) velocity difference between the northern
and the southern part of the ring, the southern part being slightly
more red-shifted. However, the ring is clearly not expanding.

A map of the 850$\,\mu$m dust continuum from \citet{sw2001} shows
faint extended emission towards V1515\,Cyg and also at about 10$''$ to
the north and northwest to the star. Although these observations were
made with a relatively large (15$''$) beam, the results are not
inconsistent with a potentially arc-shaped dust emission. It is thus
possible that the ring of material that is responsible for the optical
reflection nebula and for the $^{13}$CO emission also emits in dust
continuum.

Although the observed molecular emission does not exactly peak towards
the star, some dust and gas must be located in the vicinity of V1515
Cyg forming a circumstellar envelope. This claim is suggested by the
SED of the FUor, which was modeled with a disk+envelope geometry by
\citet{turner1997} and \citet{green2006}. However, this envelope does
not show up as a separate localized peak in our $^{13}$CO map. We
measured the highest gas mass towards this object within our sample,
which can probably be explained by the contribution from both the
circumstellar envelope and from the arc-shaped feature. Thus the
0.43\,M$_{\odot}$ is an upper limit for the envelope mass.

\subsection{V1735\,Cyg}

V1735\,Cyg had its outburst some time in the 1950s or the 1960s
\citep{elias1978}. The star is associated with faint patches and
filaments of reflected light within 1$'$. Submillimeter dust continuum
observations by \citet{sw2001} indicate the presence of two sources:
one associated with (albeit slightly offset from the optical position
of) V1735\,Cyg, and a brighter one located about 20-24$''$ to the
northeast (the deeply embedded Class\,I protostar
\object{V1735\,Cyg\,SM1}). Single-dish $^{12}$CO and $^{13}$CO
observations \citep{levreault1983,richardson1985,evans1994} revealed a
complex molecular gas structure in the vicinity of these sources. The
$^{12}$CO emission is extended in the arcminute spatial scale, and
broad line wings indicate outflow activity. \citet{evans1994}
concluded that both V1735\,Cyg and SM1 drive molecular
outflows.

Similarly to the other three FUors in our sample, our $^{13}$CO
interferometric observations of V1735\,Cyg show narrow emission with
no line wings. This again points to quiescent gas in some kind of
envelope. Our map shows a compact peak close to V1735\,Cyg, but with a
definite offset of 2$\farcs$9 to the northwest. Based on multi-filter
optical images, \citet{goodrich1987} discovered that V1735\,Cyg itself
is much more reddened than the surrounding reflection nebulosity. He
speculates that the reason for this may be a dark cloud in front of
V1735\,Cyg. In this scenario, the dark cloud may be offset from the
star, as long as it causes enough interstellar extinction to explain
the observed colors. Based on the fact that the $^{13}$CO emission we
detect is offset from the stellar position, we propose that a
significant part of the emission is coming from this foreground dark
cloud. This suggestion is supported by the results of
\citet{quanz2007}, who, based on the analysis of mid-infrared ice
features, concluded that the extinction for V1735\,Cyg might be caused
by ices somewhere in the line of sight to the source, rather than
material related to the young star. The foreground structure, however,
may not be completely unrelated to the star because of their identical
radial velocities.

For continuum observations such as those presented in
\citet{weintraub1991} and in \citet{sw2001}, the dominant source is
not the FUor, but V1735\,Cyg\,SM1. This source also emits molecular
line emission, as evidenced by the single-dish $^{12}$CO(3--2) map
presented by \citet{evans1994}. Interestingly, we did not detect
V1735\,Cyg\,SM1 in $^{13}$CO. The reason for the non-detection may be
partly due to the fact that V1735\,Cyg\,SM1 is separated from
V1735\,Cyg by about 20$''$, thus it is at the very edge of our primary
beam. Additionally, the submillimeter source is very extended, thus
the interferometer may filter out most of its emission.

\section{Summary and Conclusions}

In this paper, we present interferometric observations of the
$^{13}$CO(1--0) line of four well known FUors, young stellar objects
characterized by large optical outbursts due to enhanced disk
accretion. For V1057\,Cyg, V1515\,Cyg, and V1735\,Cyg, these represent
the first millimeter interferometric data published so far. This makes
it possible to study the gas distribution on a few thousand AU spatial
scale.

Although all of our sources are known to drive molecular outflows, as
evidenced by the high-velocity line wings of $^{12}$CO, our data
suggest that the $^{13}$CO(1--0) emission traces quiescent gas. With
the exception of V1515\,Cyg, the size of the emitting region is within
a few thousand AU, consistent with typical circumstellar envelope
sizes. Gas masses calculated from our $^{13}$CO line fluxes and from
dust continuum data from \citet{sw2001} are also consistent. This
indicates that the $^{13}$CO emission seen towards our sources is
mostly originated from a relatively compact circumstellar envelope,
or, as in the case of V1735\,Cyg, possibly from a small foreground
cloud.

All of our sources are surrounded by reflection nebulosities, which
are probably pre-existent structures illuminated by the brightened
central source. With the exception of V1515\,Cyg, we detect no
$^{13}$CO emission from these structures. For V1515\,Cyg, the
$^{13}$CO emission coincides with the ring-shaped optical reflection
nebula. This indicates that the source is surrounded by a ring of
material that on one hand scatters the optical light of the central
star, and on the other hand emits at millimeter wavelengths. An
important consequence is that attributing unresolved, single-dish
millimeter fluxes of V1515\,Cyg to the circumstellar envelope may not
be entirely correct.

Based on the appearance of the 10$\,\mu$m silicate feature,
\citet{quanz2007} defined two categories of FUors. They argue that
objects showing the feature in absorption (e.g.~V1735\,Cyg) are
younger, still embedded in a circumstellar envelope. Objects showing
the silicate band in emission (e.g.~V1057\,Cyg, V1331\,Cyg, and
V1515\,Cyg) are more evolved, with direct view on the surface layer of
the accretion disk. However, in case of V1735\,Cyg, the millimeter
emission and the absorption at shorter wavelengths may be due to a
foreground cloud, and consequently the object may be more evolved
(less embedded) than it appears. Our $^{13}$CO data, especially the
case of V1515\,Cyg and V1735\,Cyg, demonstrate that millimeter
emission and consequently absorption at other wavelengths may not be
necessarily or exclusively associated with circumstellar
envelopes. Thus, large beam, single-dish data alone are probably not
enough to obtain a complete picture of the circumstellar environment
of FUors. We suggest that any theory of the FUor phenomenon that
interprets the geometry of the circumstellar material and its
evolution using single beam measurements must be checked and compared
to interferometric observations in the future. Millimeter observations
with facilities such as ALMA for both the dust continuum and for the
line emission of molecules such as $^{13}$CO, C$^{18}$O, HCO$^+$, HCN,
etc.~will be a fruitful direction in future studies of FUors.

\begin{acknowledgements}
  The author thanks Prof.~Wolfgang J.~Duschl for making available the
  PdBI data and Dr.~Roberto Neri for his help during the data
  reduction. Discussions with Dr.~Michiel Hogerheijde and Dr.~Maria
  Kun greatly improved the presentation of the data analysis. Funding
  for the Sloan Digital Sky Survey (SDSS) and SDSS-II has been
  provided by the Alfred P.~Sloan Foundation, the Participating
  Institutions, the National Science Foundation, the U.S.~Department
  of Energy, the National Aeronautics and Space Administration, the
  Japanese Monbukagakusho, and the Max Planck Society, and the Higher
  Education Funding Council for England. The SDSS Web site is
  http://www.sdss.org/. The SDSS is managed by the Astrophysical
  Research Consortium (ARC) for the Participating Institutions. The
  Participating Institutions are the American Museum of Natural
  History, Astrophysical Institute Potsdam, University of Basel,
  University of Cambridge, Case Western Reserve University, The
  University of Chicago, Drexel University, Fermilab, the Institute
  for Advanced Study, the Japan Participation Group, The Johns Hopkins
  University, the Joint Institute for Nuclear Astrophysics, the Kavli
  Institute for Particle Astrophysics and Cosmology, the Korean
  Scientist Group, the Chinese Academy of Sciences (LAMOST), Los
  Alamos National Laboratory, the Max-Planck-Institute for Astronomy
  (MPIA), the Max-Planck-Institute for Astrophysics (MPA), New Mexico
  State University, Ohio State University, University of Pittsburgh,
  University of Portsmouth, Princeton University, the United States
  Naval Observatory, and the University of Washington.

\end{acknowledgements}

\bibliographystyle{aa}
\bibliography{paper}{}

\end{document}